\documentclass[twocolumn]{aastex63}

\turnoffedit

\usepackage{amsmath}

\graphicspath{{./}{figures/}}

\newcommand{\J}{\mathrm{SDSS\,J1137+4054}}

\begin{document}

\title{An Outburst by AM CVn binary SDSS J113732.32+405458.3}

\author[0000-0001-9195-7390]{Tin Long Sunny Wong}
\affiliation{Department of Physics, University of California, Santa Barbara, CA 93106, USA}

\author[0000-0002-2626-2872]{Jan van~Roestel}
\affiliation{Division of Physics, Mathematics and Astronomy, California Institute of Technology, Pasadena, CA 91125, USA}

\author[0000-0002-6540-1484]{Thomas Kupfer}
\affiliation{Department of Physics and Astronomy, Texas Tech University, PO Box 41051, Lubbock, TX 79409, USA}

\author{Lars Bildsten}
\affiliation{Department of Physics, University of California, Santa Barbara, CA 93106, USA}
\affiliation{Kavli Institute for Theoretical Physics, University of California, Santa Barbara, CA 93106, USA}

\correspondingauthor{Tin Long Sunny Wong}
\email{tinlongsunny@ucsb.edu}

\begin{abstract}

We report the discovery of a one magnitude increase in the optical brightness of the 59.63 minute orbital period AM CVn binary SDSS J113732.32+405458.3. Public $g$, $r$, and $i$ band data from the Zwicky Transient Facility (ZTF) exhibit a decline over a 300 day period, while a few data points from commissioning show that the peak was likely seen. Such an outburst is likely due to a change in the state of the accretion disk, making this the longest period AM CVn binary to reveal an unstable accretion disk. The object is now back to its previously observed (by SDSS and PS-1) quiescent brightness that is likely set by the accreting white dwarf. Prior observations of this object also imply that the recurrence times for such outbursts are likely more than 12 years. 

\end{abstract}

\section{Introduction}
\label{sec:intro}

AM CVn systems are hydrogen-deficient binaries with orbital periods ranging from 5 to 68 minutes \citep[e.g.,][]{Ramsay2018}. In the range $ \approx 20 - 50 $ min, they undergo outbursts similar to dwarf novae, during which the accretion disk changes from a cool low-state to a hot high-state. Both theoretical and observational studies have been carried out to understand the recurrence time, outburst duration and outburst amplitude, and probe the physics of hydrogen-deficient accretion disks \citep[e.g.,][]{Kotko2012,Levitan2015,Cannizzo2015,2018ApJ...857...52C,Cannizzo2019}. 

SDSS J113732.32+405458.3 (SDSS J1137+4054) was first identified as an AM CVn system in the Sloan Digital Sky Survey \citep[SDSS;][]{2000AJ....120.1579Y} spectroscopic data base by \cite{Carter2014}. With an orbital period of $59.63 \pm 2.74$ min measured from the radial velocities of He I emission lines, this system showed no outburst in the 9 years coverage of the Catalina Real-Time Transient Survey \citep{2009ApJ...696..870D}. \cite{Carter2014} also noted that SDSS J1137+4054 has a $u-g$ color that is redder than usual and, on a $u - g, g - r$ color-color diagram, appeared closer to DA white dwarf tracks than DB tracks. This, together with the fact that SDSS J1137+4054 has a $g - r$ color $\approx 0.1$ redder in SDSS than in the Panoramic Survey Telescope and Rapid Response System \citep[Pan-STARRS;][]{2016arXiv161205560C}, prompted us to search for variability in SDSS J1137+4054 in Zwicky Transient Facility (ZTF) public data \citep[Data Release 3;][]{2019PASP..131a8003M}. 
We also searched the Digital Access to a Sky Century @ Harvard \citep[DASCH;][]{2009ASPC..410..101G} database for lightcurves of SDSS J1137+4054. While only limiting magnitudes were available from the online database, we found an object at the location of SDSS J1137+4054 on 1937.0938 and 1939.0419 through a visual inspection of the plate images; whether this was a previous outburst is unclear.

\section{ Outburst of J1137+4054 }
\label{sec:outburst}

\begin{figure*}
\fig{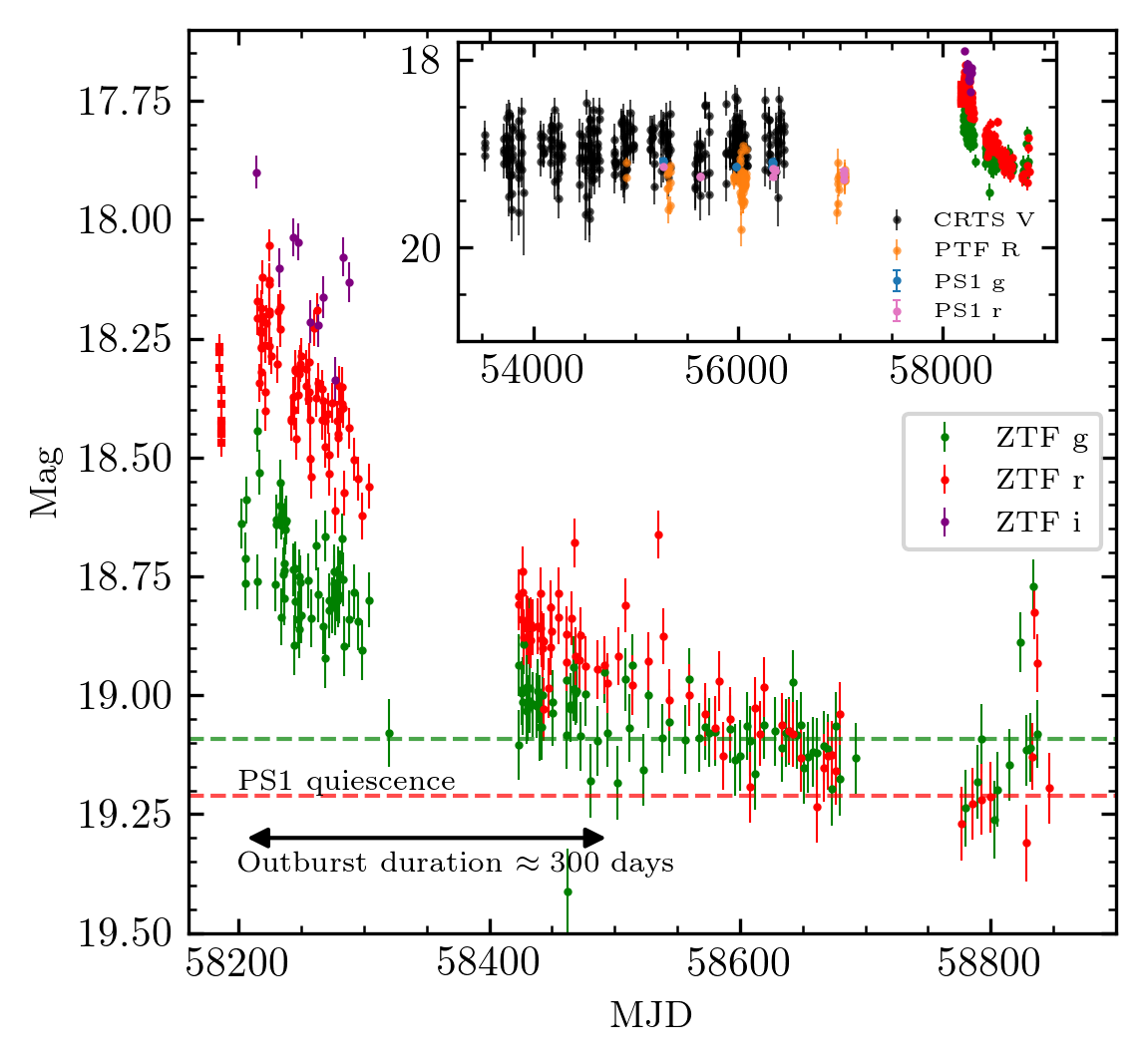}{ \textwidth}{}
\caption{ Photometry of SDSS J1137+4054 in $g$, $r$ and $i$ bands retrieved from ZTF commissioning (square) and public data (circle) and also (side panel) from CRTS ($V$ band), Pan-STARRS ($g$ and $r$ bands) and PTF ($R$ band). We also show the mean $g$ and $r$ band magnitudes from Pan-STARRS in the main panel (dashed lines), and define the outburst duration (black arrow) as the time from outburst peak to quiescence (quiescence given by $g$ band magnitude from Pan-STARRS). 
\label{fig:lightcurve}
}
\end{figure*}

We present lightcurves ($g$, $r$ and $i$ bands) of SDSS J1137+4054 obtained from ZTF public data from 2018 March 23 to 2020 January 2 in Figure \ref{fig:lightcurve}, which shows that SDSS J1137+4054 experienced an outburst starting from before 2018 March 23. We also include a few observations on 2018 March 7-9 from ZTF commissioning data, which suggests we are observing the peak of the outburst. We estimate a peak outburst magnitude in $g$ band of $\Delta m_{g} \approx 0.5$ (taking the $g$ band magnitude in Pan-STARRS as the quiescent magnitude) and estimate an outburst duration of $\approx 300$ days (time from peak to quiescence in $g$ band). We also show observations of SDSS J1137+4054 from CRTS, Pan-STARRS and the Palomar Transient Factory \citep[PTF;][]{2009PASP..121.1395L} starting from 2005 December 8, which suggest that in the 12 years prior to this outburst, SDSS J1137+4054 did not experience any outburst with a similar year-long duration. 

\section{Conclusion}
\label{sec:conclusion}

In this Research Note we report the discovery from ZTF public data of an outburst in J1137+4054. 
The outburst lasted $\approx 300$ days with an outburst magnitude in the $g$-band of $\Delta m_{g} \approx 0.5$, similar to that of SDSS J0807+4852 \citep[duration $\approx 390$ days and magnitude $\approx 2.7$;][]{Rivera2020}. 
Both events originate from long-period systems (53.3 min for J0807 and 59.6 min for J1137), 
and are much longer than predicted by theoretical calculations from the disk instability model \citep{Cannizzo2019}, which may point to enhanced mass transfer rates from the donor due to irradiation \citep[][]{Kotko2012,Rivera2020}. 


\acknowledgements
Based on observations obtained with the Samuel Oschin 48-inch Telescope at the Palomar Observatory as part of the Zwicky Transient Facility project. ZTF is supported by the National Science Foundation under Grant No. AST-1440341 and a collaboration including Caltech, IPAC, the Weizmann Institute for Science, the Oskar Klein Center at Stockholm University, the University of Maryland, the University of Washington, Deutsches Elektronen-Synchrotron and Humboldt University, Los Alamos National Laboratories, the TANGO Consortium of Taiwan, the University of Wisconsin at Milwaukee, and Lawrence Berkeley National Laboratories. Operations are conducted by COO, IPAC, and UW. 

This research was funded by the Gordon and Betty Moore Foundation through Grant GBMF5076. This research was supported in part by the National Science Foundation under Grant No. NSF PHY-1748958. 

While writing this manuscript we became aware that Rivera Sandoval et al. have written a paper reporting observations of the same outburst by SDSS J1137+4054.

\added{%
\software{%
\texttt{ipython/jupyter} \citep{perez_2007_aa,kluyver_2016_aa},
\texttt{matplotlib} \citep{hunter_2007_aa},
\texttt{NumPy} \citep{der_walt_2011_aa}, and
\texttt{Python} from \href{https://www.python.org}{python.org}.}}

\vspace{5mm}




\end{document}